\documentstyle[preprint,aps,eqsecnum,epsf] {revtex}
\begin{document}
\draft
\preprint{NCL95-TP6}
\title{Bose-Einstein condensation as symmetry breaking in compact  curved
spacetimes}
\author{John. D. Smith and David. J. Toms}
\address{Physics Department, University of Newcastle upon Tyne, NE1 7RU, United
Kingdom}
\date{\today}
\maketitle
\begin{abstract}
We examine Bose-Einstein condensation as a form of symmetry breaking in the 
specific
model of the Einstein static universe. We show that symmetry breaking never 
occurs
in the sense that the chemical potential $\mu$ never reaches its critical value.
This leads us to some statements about spaces of finite volume in general. In an
appendix we clarify the relationship between the standard statistical mechanical
approaches and the field theory method using zeta functions.
\end{abstract}

\section{Introduction}

Bose-Einstein condensation (BEC) for non-relativistic spin-0 particles is 
standard
textbook material \cite{LL,PathriaSM,Huang}. In the infinite volume limit, 
there
is a critical temperature at which a phase transition occurs. For a real system,
such as liquid helium, the effects of interactions may be important.(See
\cite{Huang93} for a recent review.) The study of Bose-Einstein condensation for
relativistic bosons is more recent. In particular, 
Refs.\onlinecite{HWPRL,HWPRD,KapPRD}
 applied the methods of relativistic quantum field theory at finite
temperature and density to study BEC. The phase transition, which occurs at high
temperatures, can be interpreted as spontaneous symmetry breaking. Subsequent
work \cite{BD,BBD} extended the analysis to self-interactions in scalar field
theory.

The generalization from flat Minkowski spacetime to curved spacetime has also
been considered. The non-relativistic Bose gas in the Einstein static universe
was given by Al'taie \cite{Alt}. The extension to relativistic scalar fields was
given for conformal coupling in Ref.\cite{PathriaSingh} and for minimal coupling
in Ref.\cite{ParkerZhang}. The higher dimensional version of the Einstein static
universe was studied by Shiraishi \cite{Shiraishi}. More recently, the case of
hyperbolic universes \cite{CV} and the Taub universe \cite{WHuang} have received
attention. Anti-de~Sitter space was studied in Ref.\cite{VanzoTurco} where some 
of the issues of our paper where considered from a different viewpoint.

One advantage of dealing with specific spacetimes of the type mentioned above is
that the eigenvalues of the Laplacian are known, and as a consequence the
partition function and the thermodynamic potential can be obtained in closed
form. Another approach to studying BEC is to try to keep the spacetime fairly
general, and to calculate the thermodynamic potential only in the high
temperature limit. This has been done by a variety of people 
\onlinecite{DK,DSPRD,DSNP,KirstenJPA,KirstenCQG,TomsPRL,TomsPRD}. In particular 
the
symmetry breaking interpretation of BEC was given in Refs. 
\cite{TomsPRL,TomsPRD}.
The effects of interactions have been given recently \cite{KT}.

The purpose of the present paper is to re-examine BEC in the case where the
spatial manifold is compact. We will be particularly concerned with the Einstein
static universe for which the spatial manifold is $S^3$. Because the volume is
finite the general theory presented in Refs. \cite{TomsPRL,TomsPRD} must be
modified since the results tacitly assumed the infinite volume limit. Although
the thermodynamics of finite volume systems has been well studied in classical
nonrelativistic statistical mechanics 
\cite{Gob1,Gob2,Gob3,PathPR,GreenPath1,GreenPath2,GreenPath3,PathZasa1,PathZasa2},
 there appears to
be some confusion in the literature in the context of relativistic field theory,
particularly in the calculation of critical temperatures. After a brief review 
of
the general theory presented in Refs.\cite{TomsPRL,TomsPRD}, we will specialize
to the Einstein static universe. The generalised $\zeta$~-function 
\cite{Hawking}
will be evaluated for this spacetime and used to obtain the effective action for
a complex scalar field with a finite charge. Expressions for the pressure and
charge are obtained. We then present a detailed analysis of whether BEC and
symmetry breaking occur, and conclude that they cannot, in contrast to the
expectations of Refs.\cite{TomsPRL,TomsPRD}. The reason for the difference is
linked to the fact that the spatial manifold has a finite volume, and we present
a discussion of this important point. In appendix \ref{appendix1} we show how 
the
generalised $\zeta$~-function can be used to relate the thermodynamic potential
directly to the effective action. In appendix \ref{appendix2} we present a short
discussion of how the analysis proceeds for the anisotropic spacetime obtained 
by
identification of antipodal points on $S^3$.

\section{The Effective Action and BEC}
\label{sec:two}

We will consider a complex scalar field defined on a $(D+1)$-dimensional
spacetime manifold $M \simeq R \times \Sigma$ where $\Sigma$ is a compact
Riemannian manifold of dimension $D$. After a Wick rotation to imaginary time 
the
configuration space action is 
\begin{eqnarray}
\tilde{S}[\varphi] &=& \int_0^{\beta} dt \int_{\Sigma} d\sigma_x \biggl\{ 
\frac{1}{2}
(\dot{\varphi}_1 - i \mu\varphi_2)^2 + \frac{1}{2}(\dot{\varphi}_2 +
i\mu\varphi_1)^2\nonumber \\ 
& & + \frac{1}{2}\vert \nabla\varphi_1 \vert^2 +
\frac{1}{2}\vert\nabla\varphi_2\vert^2 + \frac{1}{2} (m^2 + \xi R)(\varphi_1^2 +
\varphi_2^2) \biggr\} .\label{2.1}
\end{eqnarray} (Here $\vert \nabla \varphi_1 \vert^2 = g^{ij}\partial_i 
\varphi_1
\partial_j \varphi_1$ with $g_{ij}$ the Reimannian metric on $\Sigma$.
$d\sigma_x$ is the invariant volume element on $\Sigma$.) This may be found as
described by Kapusta \cite{KapPRD} in flat spacetime starting with the phase
space path integral and incorporating the conserved charge using a Lagrange
multiplier $\mu$. The partition function may be expressed as a path integral 
over
all fields periodic in time with period $\beta = T^{-1}$ where $T$ is the
temperature.

It is straightforward to compute the effective action using the background field
method \cite{DeWitt65}. If we take the background field to be 
$\bar{\varphi}_1(x)
= \bar{\varphi}(x)$, $\bar{\varphi}_2(x) = 0$, then the effective action is 
\begin{equation}
\Gamma [\bar{\varphi}] = \tilde{S}[\bar{\varphi}] + \frac{1}{2} \ln \det \{l^2
\tilde{S}_{,ij}[\varphi]  \} \label{2.2}
\end{equation} using condensed notation \cite{DeWitt65}. $l$ is a unit of 
length introduced to keep the argument of the logarithm dimensionless.

Let $\{ \varphi_N(x) \}$ be a complete set of solutions to
\begin{equation} [- \nabla^2 + \xi R]\varphi_N(x) = \sigma_N\varphi_N(x)
\label{2.3}
\end{equation} normalized by
\begin{equation}
\int_\Sigma d\sigma_x \varphi_N(x)\varphi_{N'}(x) = \delta_{NN'} \label{2.4}
\end{equation} If we define a generalized $\zeta$-function by
\begin{equation}
\zeta_{\cal D}(s) = \sum_{j = -\infty}^{\infty} \sum_N \left[ \left({2\pi j\over
\beta} + i\mu \right)^2 + \sigma_N + m^2\right]^{-s}, \label{2.5}
\end{equation} then it can be shown that \cite{TomsPRD} 
\begin{equation}
\Gamma [\bar{\varphi}] = \tilde{S}[\bar{\varphi}] - \zeta_{\cal D}'(0) +
\zeta_{\cal D}(0)\ln l^2 . \label{2.6}
\end{equation}

In order for the sum in (\ref{2.5}) to be well defined, $\mu$ must be restricted
by $\sigma_0 + m^2 - \mu^2 > 0$, where $\sigma_0$ is the smallest eigenvalue of
the set.

The background scalar field must satisfy
\begin{equation} {\delta \Gamma [\bar{\varphi}] \over \delta
\bar{\varphi}(x)} = 0. \label{2.7}
\end{equation} The differentiation is to be computed with $\mu$, $T$, $V$,
$g_{ij}$ all held fixed. This is equivalent to minimizing the Helmholtz free
energy with $Q$, rather than $\mu$, held fixed \cite{Toms94}. Because
$\zeta_{\cal D}$ has no explicit dependence on $\bar{\varphi}$, using 
(\ref{2.6}) 
in (\ref{2.7}) gives
\begin{equation} - \nabla^2 \bar{\varphi} + \left( m^2 - \mu^2 + \xi R  \right)
\bar{\varphi} = 0. \label{2.8}
\end{equation} We can expand $\bar{\varphi}$ in terms of the $\varphi_N$ , which
satisfy (\ref{2.3},\ref{2.4}) :
\begin{equation}
\bar{\varphi}(x) = \sum_N C_N \varphi_N(x). \label{2.9}
\end{equation} where the expansion coefficients $C_N$ are to be determined. Use
of (\ref{2.9}) in (\ref{2.8}) leads to 
\begin{equation} (\sigma_N + m^2 - \mu^2) C_N = 0. \label{2.10}
\end{equation} If $\mu^2 < \sigma_0 + m^2$ then the only possible solution to
(\ref{2.10}) is $C_N = 0$ for all $N$. This leads to $\bar{\varphi}(x) = 0$ as
the ground state and no symmetry breaking. If, however, it is possible for $\mu$
to reach the critical value $\mu_c$ defined by
\begin{equation}
\mu_c^2 = \sigma_0 + m^2 \label{2.11}
\end{equation} then $C_0$ in (\ref{2.10}) is undetermined. In this case the
ground state in (\ref{2.9}) is
\begin{equation}
\bar{\varphi}(x) = C_0 \varphi_0(x) , \label{2.12}
\end{equation} and there is symmetry breaking.

In order to find the relation with BEC, consider the charge which is given by (
in units with $e=1$)
\begin{equation} Q =  - \frac{1}{\beta} {\partial \Gamma \over \partial \mu} .
\label{2.13}
\end{equation} From (\ref{2.6}) it is clear that we can write
\begin{equation} Q = Q_0 + Q_1  \label{2.14}
\end{equation} where
\begin{equation} Q_0 = \mu \int_\Sigma d\sigma_x\bar{\varphi}^2(x) ,  
\label{2.15}
\end{equation}
\begin{equation} Q_1 =  \frac{\partial}{\partial\mu} \{\zeta_{\cal D}'(0) -
\zeta_{\cal D}(0)\ln  l^2 \}.  \label{2.16}
\end{equation} Use of the high temperature expansions in Ref.\cite{TomsPRD} 
shows
that for T large enough, it is always possible to have $\mu < \mu_c$ and
$\bar{\varphi} = 0$. In this case $Q_0 = 0$. When the temperature drops, $\mu$
increases. If it is possible for $\mu$ to reach the value $\mu_c$ defined in
(\ref{2.11}), $\bar{\varphi} \neq 0$ and $Q_0 \neq 0$. The value of $T$ at which
$\mu = \mu_c$ is defined to be the critical temperature $T_c$. It was shown in
Ref.\cite{TomsPRD} that
\begin{equation} T_c \approx \left( {3 Q \over \mu_c V} \right)^{1 \over 2} 
\label{2.17}
\end{equation} exactly as in flat spacetime \onlinecite{HWPRD,KapPRD} (allowing
for a different value of $\mu_c$). For $T < T_c$, we have
\begin{equation} Q_0 = Q \left[ 1 - \left({T \over T_c}\right)^2 \right] ,
\label{2.18}
\end{equation}
\begin{equation} C_0 = \left({V \over 3}\right)^{1 \over 2} \left( T_c^2 - T^2
\right)^{1 \over 2} .  \label{2.19}
\end{equation}

These results assume that it is possible for $\mu$ to reach the critical value
$\mu_c$ at a finite temperature. However without a detailed analysis it is
difficult to see if this is possible. For example, in the non-relativistic case,
when $\Sigma$ is a flat $2$ - dimensional space, $\mu$ never reaches $\mu_c$ and
therefore there is no BEC. ( See Ref.\cite{May59} for example.) Another case
where $\mu$ never reaches $\mu_c$ is for $\Sigma$ a flat $3$ - dimensional space
with an externally applied constant magnetic field \cite{Schafroth55}. More
generally, if $\Sigma$ has a finite volume extreme care must be used. In the
past, compact spaces have been analyzed using high temperature expansions. A
proper analysis of BEC may require temperatures outside the range where the high
temperature expansion is valid. Use of the high temperature expansion may give
misleading results as we show later. A recent discussion of the existence of BEC
using generalised $\zeta$~-functions has recently been given \cite{DandK}.

\section{The Einstein Universe}
\label{sec:one}
 
At this point it is instructive to consider a specific example. One space which 
has received considerable attention in the past is the Einstein static universe, 
ie $\Sigma = S^3$ which has radius $a$, with scalar curvature $R = 6a^{-2}$. We 
will take $U_1 = \xi R$, where $\xi$ must satisfy the condition $\xi > 
\frac{1}{6}(1 - a^2 m^2)$ but is otherwise arbitrary. High temperature 
expansions for the case $\xi = 0$ where first obtained by Parker and Zhang 
\cite{ParkerZhang}, whilest the case $\xi = \frac{1}{6}$ has been studied by 
Singh and Pathria \cite{PathriaSingh}. Previously the critical temperature has 
been calculated to be 
\begin{equation} T_c = \left[{3q \over m}\right]^{1 \over 2}\left[1 + {6\xi 
\over 
m^2a^2}\right]^{- {1 \over 4}}
\end{equation} (see \cite{TomsPRD} and \cite{ParkerZhang} for the case $\xi = 
0$). $q = Q/V = Q/2\pi^2 a^3$ is the charge density.

It is well known \onlinecite{Schrod,Erdelyi,Ford75}
that the eigenvalues of $-\nabla^2$ on $S^3$ are $-N(N + 
2)a^{-2}$, with a degeneracy of $(N + 1)^2$, ($N = 0, 1, 2, \ldots$). Hence
$\sigma_N = [ N ( N + 2) + 6\xi ]/ a^2$ and our generalised zeta function is 
\begin{equation}
\zeta_{\cal D}(s) = \sum_{j = -\infty}^{\infty} \sum_{N = 0}^{\infty} {(N + 1)^2
\over [({2\pi \over \beta}j + i\mu)^2 + {N (N + 2) + 6\xi \over a^2} + m^2]^s}
\label{3.1}
\end{equation}

A slight simplification
in the form of (\ref{3.1}) is obtained by completing the square in $N$ in the
denominator and then relabelling the sum, $n = N + 1$. Thus we consider the
analytic continuation of
\begin{equation}
\zeta(s) = \sum_{j = -\infty}^{\infty}\sum_{n = 1}^{\infty}{n^2 \over [(aj + 
ib)^2 + \alpha n^2 + c]^s}, \label{3.2}
\end{equation}
where $a = 2\pi / \beta$, $b = \mu$, $\alpha = 1/ a^2$ and~$c = m^2 + (6\xi -
1)/ a^2$. The techniques which we shall use are based on those of Elizalde
\cite{Elizalde1,Elizalde2,Elizalde3}.

We start by making use of the identity
\begin{equation} 
a^{-s} = \frac{1}{\Gamma (s)}\int_0^{\infty} dt\ t^{s-1}\exp (-at) ,
\label{3.3}
\end{equation}
and, expanding out the $(aj + ib)^2$ to get 
\begin{equation}
\zeta(s) = {1 \over \Gamma(s)} \int_0^{\infty}dt\ t^{s-1} \sum_{n=1}^{\infty}
n^2 \exp (-\alpha n^2t) \\
\sum_{j=-\infty}^{\infty} \exp (-aj^2t - 2iajbt - ct + b^2 t) .
\label{3.4}
\end{equation}
(We shall justify interchanging the sums with integrals shortly).
Now, following the notation of Whittaker and Watson \cite{WW} for theta
functions,
\begin{eqnarray}
\sum_{j=-\infty}^{\infty} e^{-aj^2t - 2ajbti} &=& 1 + 2\sum_{j=1}^{\infty} 
e^{-a^2j^2t}\cos (2abjt) \nonumber \\
&=&\theta_3\left(abt \left| i{a^2t \over \pi}\right.\right)
\end{eqnarray}
where $\theta_3(z|\tau)$ is defined as ($q = e^{\pi i\tau}$)
\begin{equation}
\theta_3(z | \tau) = \theta_3(z, q) = 1 + 2\sum_{n=1}^{\infty} q^{n^2} \cos 2nz 
.
\end{equation}
One immediately notes that
\begin{equation}
\sum_{n=1}^{\infty} n^2 e^{-\alpha n^2 t} = - {\partial \over \partial (\alpha 
t)} \theta_3 \left(0 \left| i{\alpha t \over \pi}\right.\right) .
\end{equation}

An important property  of the theta functions is that they are uniformly 
convergent for ${\cal I}m\  \tau > 0$, as is their derivative. Thus the 
integrand 
in (\ref{3.4}) is the product of a number multiplied by two uniformly convergent 
sums. This justifies our previous manipulations.

We are now in a position to rewrite our zeta function in terms of theta 
functions,
\begin{equation}
\zeta(s) = \frac{1}{\Gamma (s)} \int_0^{\infty} dt\ t^{s-1}  
e^{-(c-b^2)t}\left(-{\partial
\over \partial (\alpha t)} 
\theta_3\left(0\left|i{\alpha t \over \pi}\right.\right)\right) 
\theta_3\left(abt\left|i{a^2t 
\over \pi}\right.\right) 
\end{equation}
Of course, although apparently much simpler, the integral as it stands could 
only be performed numerically since there is no representation for the theta 
function, other than in terms of the infinite sum above. However progress can 
be made by making use of Jacobi's imaginary transformation for theta functions:
\begin{equation}
\theta_3(z|\tau) = (-i\tau)^{-{1 \over 2}} \exp \left({z^2 \over \pi 
i\tau}\right)\theta_3\left({z \over \tau}\left|- {1 \over \tau} \right.\right) 
.
\label{3.5} 
\end{equation} 

Applying this to the second theta function gives us
\begin{equation}
\zeta(s) = {1 \over \Gamma(s)} {\pi^{1 \over 2} \over a} \int_0^{\infty} dt \, 
t^{s - {3 \over 2}} e^{-ct} \sum_{n=1}^{\infty} n^2 e^{-\alpha n^2 t} \\
\left( 1 + 2 \sum_{j=1}^{\infty} e^{- {j^2\pi^2 \over a^2 t}} \cosh (2j{b \over
 a}\pi) \right) .
\end{equation}
We now split this expression into the sum of two terms as follows:
\begin{equation}
\zeta(s) = I_1 + I_2 ,
\end{equation}
 where
\begin{mathletters}
\begin{equation} I_1 = \frac{1}{\Gamma (s)}\frac{\pi^{1 \over 2}}{a}
\int_0^{\infty} dt\ t^{s -  {3 \over 2}} e^{-ct} \sum_{n=1}^{\infty} n^2 
e^{-\alpha
n^2 t}
\end{equation} 
and
\begin{equation} I_2 = {2\pi^{1 \over 2} \over a\Gamma (s)} \sum_{n=1}^{\infty}
n^2 
\sum_{j=1}^{\infty} \cosh \left({2\pi jb \over a} \right) \int_0^{\infty} dt\ 
t^{s - {3 \over 2}} \exp \Bigl[ -(c + \alpha n^2)\,t - \frac{j^2\pi^2}{a^2t} 
\Bigr]
\label{3.6}
\end{equation}
\end{mathletters} If we now differentiate (\ref{3.5}) w.r.t $\alpha t$~, we
obtain the useful  identity
\begin{equation}
\sum_{n=1}^{\infty} n^2 e^{-\alpha n^2t} = {\pi^{1 \over 2} \over 4(\alpha t)^{3 
\over 2} } + {\pi^{1 \over 2} \over 2(\alpha t)^{3 \over 2} }\sum_{n=1}^{\infty} 
\exp \Bigl({n^2\pi^2 \over \alpha t}\Bigr) - \left({\pi \over \alpha 
t}\right)^{5 \over 2} \sum_{n=1}^{\infty} n^2 \exp \Bigl({n^2\pi^2 \over \alpha 
t}\Bigr)
\label{3.7}
\end{equation} which, together with a standard integral representaion for the 
modified Bessel function \cite{Grad} 
\begin{equation}
K_{-\mu} (z) = \frac{1}{2}\left({2 \over z}\right)^\mu \int_0^{\infty} dt\, 
t^{\mu -
1} \exp \left( - t - {z^2 \over 4t} \right) ,
\end{equation}
gives us
\begin{equation} 
I_1 = {\pi c^{2 - s} \over 4 (s-1)(s-2) a \alpha^{3 \over 2} } +
{(2c)^{2 - s} \pi \over a\alpha^{3 \over 2}\Gamma (s)}\sum_{n=1}^{\infty} 
\frac{1}{X_n^{2 - s}} \{ K_{- s + 2}(X_n) - X_n K_{- s + 3}(X_n)\}\ ,
\end{equation} 
where
\begin{equation}
X_n = {2n\pi c^{1 \over 2} \over \alpha^{1 \over 2}}.
\end{equation} 
Similarly, 
\begin{equation} 
I_2 = {2\pi^{1 \over 2} \over a \Gamma (s)}\sum_{j=1}^{\infty}
\cosh \left(  {2\pi jb \over a} \right) \sum_{n=1}^{\infty} 2n^2 (c + \alpha
n^2)^{-s+{1 \over  2}} \left[ {j\pi (c + \alpha n^2)^{1 \over 2} \over a}
\right]^{s - {1 \over 2}}  K_{-s+{1 \over 2}}(Z_{jn})\ ,
\end{equation}
where
\begin{equation}
 Z_{jn} = 2\pi j{(c + \alpha n^2)^{1 \over 2} \over a}.
\end{equation} 
(Note that we have not used equation (\ref{3.7}) here). It can
be shown that the sums in $I_1$ and $I_2$ are uniformly convergent for  all $s$
(this is because $K_\mu (z)$ falls off exponentially for large $z$, 
and $K_\mu =  K_{-\mu}$). Thus we finally obtain the expansion
\begin{eqnarray}
\zeta (s) = {\pi c^{2 - s} \over 4 (s-1)(s-2) a \alpha^{3 \over 2} } +
{(2c)^{2 - s} \pi \over a\alpha^{3 \over 2}\Gamma (s)}\sum_{n=1}^{\infty} 
\frac{1}{X_n^{2 - s}} \{ K_{- s + 2}(X_n) - X_n K_{- s + 3}(X_n&&)\}\ 
\nonumber \\
 + {2\pi^{1 \over 2} \over a \Gamma (s)}\sum_{j=1}^{\infty} \cosh \left( {2\pi 
jb \over a} \right) \sum_{n=1}^{\infty} 2n^2 (c + \alpha n^2)^{-s+{1 \over 2}} 
\left[ {j\pi (c + \alpha n^2)^{1 \over 2} \over a} \right]^{s - {1 \over 2}}&& 
\nonumber \\
 \times \,K_{-s+{1 \over 2}}(Z_{jn})\ .&&
\end{eqnarray}

Now, for our application, we are only interested in $\zeta (0)$ and~$\zeta'(0)$. 
At first sight it looks like our effective action will be extremely complicated, 
however 
\begin{equation}
\frac{1}{\Gamma (s)} \sim s + \gamma s^2 + O\bigl(s^3\bigr),
\quad
\hbox{as $s \to 0$,}
\label{3.8}
\end{equation}  and so we are only left with the first term in $\zeta (0)$.
Similarly, when we  differentiate with respect to $s$ and then evaluate at 
$s=0$~,
the result is to  differentiate the first term and then simply to remove the
gamma functions from  all subsequent terms. Furthermore, using \cite{Grad}
\begin{equation} K_{1 \over 2}(z) = \left( {\pi \over 2 z} \right)^{1 \over 2}
e^{- z}
\end{equation} the double sum term in $\zeta'(0)$ becomes
\begin{mathletters}
\begin{eqnarray} 
2\sum_{j=1}^{\infty} {\cosh \left({2\pi jb \over a} \right)
\over j} 
\sum_{n=1}^{\infty} n^2 \exp \left\{ - {2\pi j (c + \alpha n^2)^{1 \over 2} 
\over a} \right\} \label{twenty} &=& \sum_{n=1}^{\infty} n^2 \sum_{j=1}^{\infty}
{e^{-j \omega_{n+} }+ e^{-j 
\omega_{n-}} \over j} \\ 
&=& - \sum_{n=1}^{\infty} n^2 \left\{ \ln \left(1 - e^{-
\omega_{n+}} \right)\right. \nonumber \\
& &\ \ \ \ \ \ \ \ \ \ \ \ \ + \left.\ln \left( 1 - e^{- \omega_{n-}} \right) 
\right\}\label{tw-one}
\end{eqnarray} 
where
\begin{equation}
\omega_{n \pm} = \frac{2 \pi}{a} \bigl(c + \alpha n^2 \bigr) \pm {2\pi b \over 
a} 
\end{equation}
\end{mathletters} 
which is the usual statistical mechanical contribution. 
(See appendix \ref{appendix1}).

Putting all of the above together, and rewriting the effective action 
in terms of the physical variables, we obtain
\begin{eqnarray}
\Gamma = && \tilde{S}[\bar{\varphi}] + \frac{1}{16} \beta a^3 c^2 \ln (l^2c) - 
\frac{3}{32} \beta a^3 c^2 \nonumber \\ 
&& - 2
\beta a^3 c^2 \sum_{n=1}^{\infty} \frac {1}{\bigl( 2 \pi na c^{1 \over 2} 
\bigr)^2 } \left\{ K_2\bigl(2 \pi na c^{1 \over 2}\bigr) - 2 \pi na c^{1 \over 
2} K_3\bigl(2 \pi na c^{1 \over 2}\bigr) \right\} \nonumber \\
&& +
\sum_{n=1}^{\infty} n^2 \ln \left[ \left(1 - e^{- \beta (\omega_n + \mu)} 
\right) \left(1 - e^{- \beta (\omega_n - \mu)} \right) \right]
\label{3.9}
\end{eqnarray} 
where $\omega_n = \bigl( (n/a)^2 + c \bigr)^{1 \over 2}$~, $c =
m^2 + (6 \xi -  1)/a^2$ and we expect $\tilde{S}[\bar{\varphi}]$ to be zero for
$T > T_c$.

In the above we have been exclusively concerned with the temperature zeta 
function. Originally it was hoped that by concentrating on the full sum one may 
be able to find some simplification which would not be apparent if the sum was 
separated into its zero temperature and finite temperature contributions. 
One approach would be try to simplify the sum over $n$ by using (\ref{3.7})
in (\ref{3.6}). Making liberal use of the summation formulas in the appendix
of reference \cite{PathriaSingh} one finally obtains
\begin{eqnarray}
\lefteqn{\Gamma = \tilde{S}[\bar{\varphi}] + \frac{a^3\beta}{16}c^2\ln (l^2c) 
 - \frac{3a^3c^2\beta}{32} } \nonumber \\
& & \mbox{} + \frac{2\pi a^3c}{\beta} \sum_{j=1}^{\infty} \frac{\cosh 
(\beta \mu j)}{j^2} K_2\bigl(\beta c^{1 \over 2} j \bigr) \nonumber \\
& & \mbox{} + \frac{1}{4\pi^2}\sum_{l=-\infty}^{\infty}\left\{ - \frac{A^2}{a} 
\ln (1 - y) + \frac{2A}{a^2} g_2(y) + \frac{2}{a^3} g_3(y) \right\}
\end{eqnarray}
where
\begin{equation}
A = 2\pi \left[ c - \bigl(\mu + i \frac{2\pi l}{\beta}\bigr)^2 \right]^{1 \over
2},
\end{equation}
\begin{equation}
y = \exp(- a A )
\end{equation}
and the $g_n(y)$ are the generalised Bose-Einstein integrals, defined by
\begin{eqnarray}
g_n(y) &=& \frac{1}{\Gamma(n)} \int_{0}^{\infty} {x^{n-1}\, dx \over y^{-1}e^x
- 1} \\
&=& \sum_{l=1}^{\infty} {y^l \over l^n}
\end{eqnarray}
(see \cite{PathriaSM} appendix D for a discussion of the properties of these 
functions).
Differentiating with respect to $\mu$ gives a conserved charge of
\begin{equation}
Q = \frac{2\pi a^3 c}{\beta} \sum_{j=1}^{\infty} \frac{\sinh (\beta \mu j)}{j}
K_2(\beta c^{1 \over 2} j)
- 2a^3 \sum_{l=-\infty}^{\infty}\frac{\bigl[c - ( \mu + i \frac{2\pi l}{\beta} 
)^2 \bigr]^{1 \over 2} }{ e^{-a A} - 1}
\end{equation}
in agreement with \cite{PathriaSingh}, where they used the Poisson-Summation 
formula to derive this result. Examination of this result shows that it is 
convergent 
for $\mu^2<c$; however (\ref{3.9}) is convergent in the larger range $\mu^2 < 
c + 1/a^2$. The reason for this discrepancy is most easily understood by 
consideration of the statistical mechanical term in (\ref{3.9}). We can then
see how the use of the Poisson-Summation formula results in a shortened 
convergence
regime.

\begin{figure}
\begin{center}
\leavevmode
\epsfxsize=15cm
\epsffile{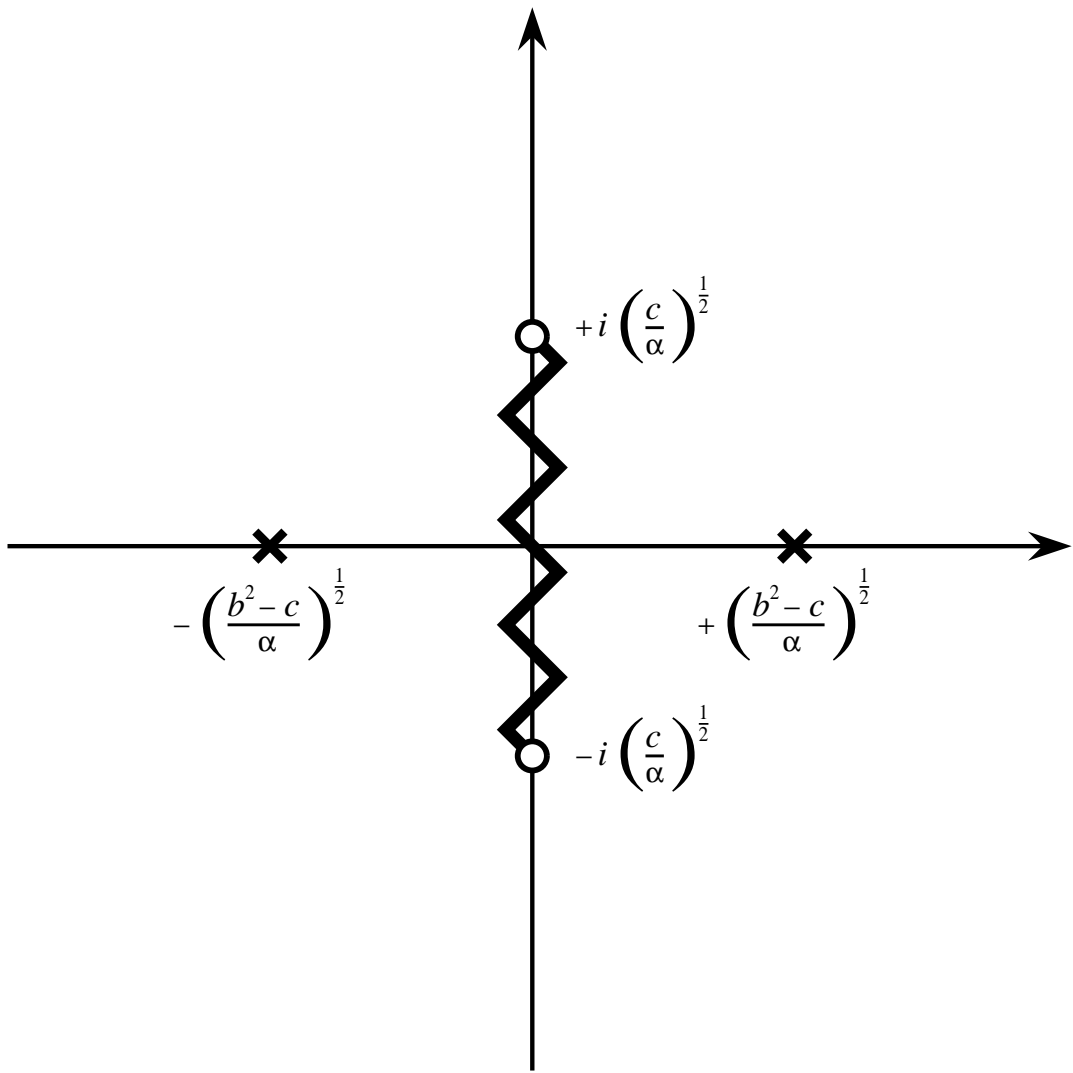}
\end{center}
\caption{Analytic structure of $f(n)$ in the complex $n$ plane, showing the 
position of branch cuts and singularities for the case $b^2 > c$.} \label{fig1}
\end{figure}

The Poisson-Summation formula essentially replaces the original sum with a new 
sum over the Fourier modes of the original. For an even summand it becomes:
\begin{equation}
\sum_{n=1}^{\infty} f(n) = \frac{1}{2} \sum_{q=-\infty}^{\infty}
\int_{-\infty}^{\infty} f(n) e^{2\pi iqn}\, dn .
\end{equation}
In our case,
\begin{equation}
f(n) = n^2 \ln \left[\left(1 - \exp -\frac{2\pi}{a}[(c + \alpha n^2)^\frac{1}{2}
- b]\right) \left(1 - \exp -\frac{2\pi}{a}[(c + \alpha n^2)^\frac{1}{2}
+ b]\right) \right]
\end{equation}
The analytic structure of $f(n)$ in the complex plane is shown in Fig.\ 
\ref{fig1}.
 There are
isolated essential singularities at $n = \pm [b^2 - c/\alpha]^{1 \over 2}$, and 
a
branch cut between $\pm i(c/\alpha)^{1 \over 2}$. When $b^2<c$ the singularities
lie on the imaginary axis and hence play no role; however for $b^2>c$ the 
singular points are real and tend to 1 as $b^2 \to c + \alpha^2$ at which point 
the original sum diverges. Use of the Poisson-Summation formula implicitly
assumes that $f(n)$ is analytic so that its Fourier integral decomposition 
exists
and hence cannot be used for $b^2 > c$. Similarly, the approach above also 
implicitly assumes that the function is analytic along the real line in the 
evaluation of the Bessel function integrals. In performing the analytic 
continuations one should be careful to ensure that the domain of convergence of
the final result is the same as that of the original zeta function.

Returning now to the full result of equation (\ref{3.9}) we can now proceed to
write down the thermodynamic potentials.
Remembering that $\Gamma = - q = - \beta PV$, and using the volume of $S^3$, $V 
= 2 \pi^2 a^3$, we can immediately write down the pressure (for $T>T_c$)
\begin{eqnarray} 
P = &-& \frac{1}{32 \pi^2} c^2 \ln (l^2c) + \frac{3 c^2}{64 \pi^2} \nonumber \\
&+& \frac{c^2}{\pi^2}
\sum_{n=1}^{\infty} \frac {1}{\bigl( 2 \pi na c^{1 \over  2} \bigr)^2 } \left\{
K_2\bigl(2 \pi na c^{1 \over 2}\bigr) - 2 \pi na c^{1 
\over 2} K_3\bigl(2 \pi na c^{1 \over 2}\bigr) \right\} \nonumber \\
&-& \frac{1}{2 \pi^2 a^3\beta} \sum_{n=1}^{\infty} n^2 \ln \left[ \left(1 - e^{- 
\beta (\omega_n + \mu)} \right) \left(1 - e^{- \beta (\omega_n - \mu)} \right) 
\right].
\label{3.10}
\end{eqnarray} 
The first few terms in the pressure are essentially unobservable
renormalisation  constants (since one can only really measure differences in
pressure), and the  last term is the term that one would predict from normal
statistical mechanics.  The Bessel function term, however, is observable since 
it
varies with volume; it  is a contribution to the pressure due to the Casimir
effect. Notice (as one would  expect) that it is independent of both the
temperature and chemical potential -  its origin is in the vacuum, not the 
finite temperature and chemical potential effects. 
Also, because of the  asymptotic form of the Bessel functions, this term
tends towards zero as the  volume (and therefore $a$) tends to infinity; it will
not survive the infinite volume limit.

Equation (\ref{3.10}) on its own does not completely determine the pressure: 
$\mu$ is unknown. To rectify this, we must now consider the vacuum expectation 
value of the charge. Above the critical temperature it is given by 
(\ref{2.13})
\begin{eqnarray} 
Q &=& \sum_{n=1}^{\infty} {n^2 \over e^{\beta (\omega_n + \mu)}- 1} +
\sum_{n=1}^{\infty} {n^2 \over e^{\beta (\omega_n - \mu)} - 1} \nonumber \\
 &=& \sum_{n=1}^{\infty} { 2n^2 \sinh (\beta \mu) e^{\beta \omega_n} \over
\bigl( e^{\beta (\omega_n + \mu)} - 1 \bigr) \bigl( e^{\beta (\omega_n - \mu)} - 
1 \bigr) }
\label{3.11}
\end{eqnarray}
 The charge $Q$ is made up from two parts: one contribution from
particles  (chemical potential $\mu$), and a contribution from antiparticles
(chemical  potential $ - \mu$). Its high temperature behaviour is examined in 
\cite{ParkerZhang}. The low temperature behaviour ($\beta \to \infty$) would be 
trivial to calculate, except that a priori one can not be sure that 
(\ref{3.11}) still holds.

\begin{figure}
\begin{center}
\leavevmode
\epsfxsize=15cm
\epsffile{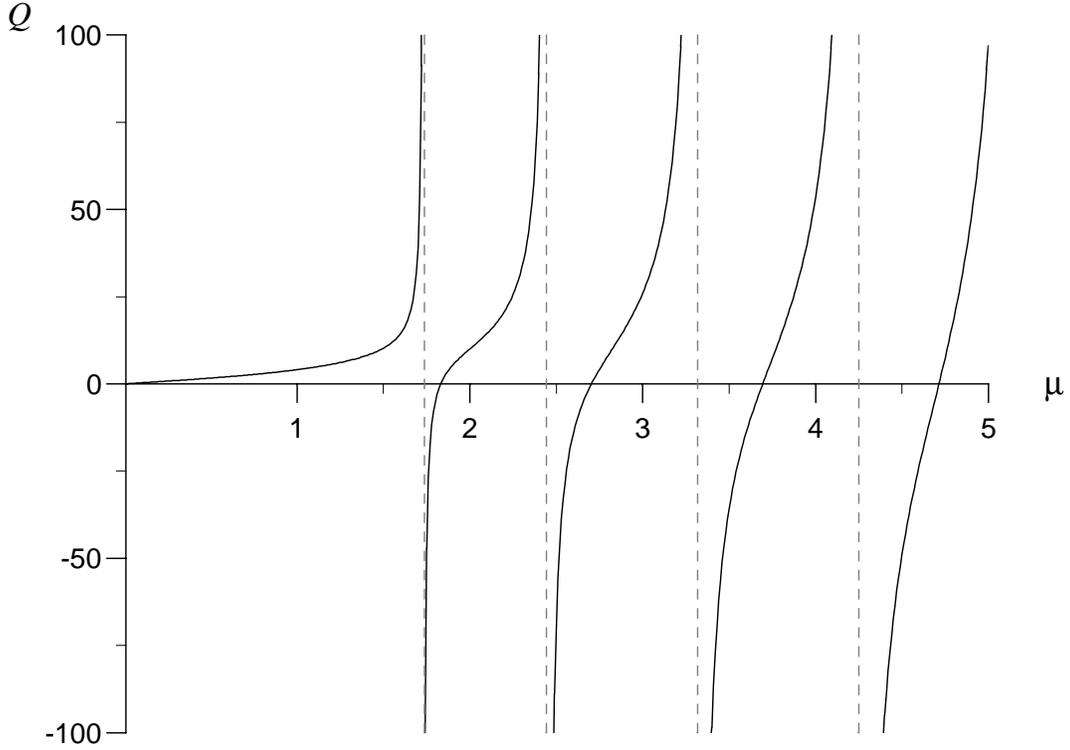}
\caption{Plot of charge $Q$ against $\mu$. } \label{charge}
\end{center}
\end{figure}

If one plots (\ref{3.11}) as a function of $\mu$ one obtains a graph similar 
to Fig.\ \ref{charge} .The charge diverges whenever $\mu$ tends to one of the 
$\omega_n$, and it is  
multivalued, being divided up into intervals by these values. Now  from
(\ref{3.11}) $\mu$ and its' derivatives are well defined within each of  these
intervals (since the sums are absolutely convergent) hence $\mu$ is  a 
continuous
function of $T$ in these regions. Furthermore, within each of these  regions $Q$
takes on all values; hence after choosing one of these regions one can never
evolve out of it by a change of $Q$ or $T$. Since a change of $Q \to  -Q$ (i.e
swapping particles with antiparticles) must cause $\mu \to - \mu$ only  the 
first
region is physical.

\begin{figure}
\begin{center}
\leavevmode
\epsfxsize=15cm
\epsffile{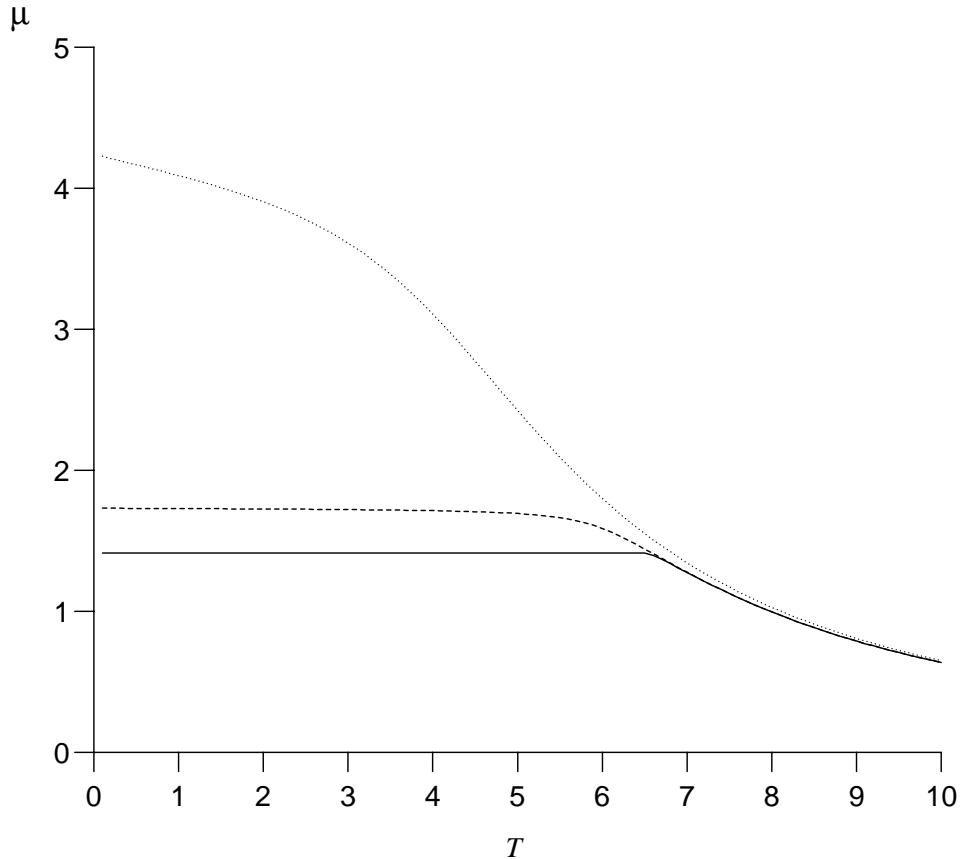}
\caption{Plots of $\mu$ against Temperature. Curves have $m^2=2$ and $\xi=1/6$.
The dotted line corresponds to $a=0.25$, the dashed line is $a=1.0$ and the 
solid
line is $a=20$.} \label{fig2}
\end{center}
\end{figure}

Although we cannot solve for $\mu$ analytically, in the case of the Einstein 
Universe we have done so numerically. The results are shown in Fig.\ \ref{fig2} 
(which uses 
the charge density $\rho = Q/V$ rather than $Q$. From the figure, one can see 
that in fact there is no critical temperature: $\mu$ tends towards its critical 
value, only reaching it at $T=0$. This is most marked in the case $a = 0.25$. 
When the volume becomes large, the curve for $\mu$ starts to look more like the 
scenario given above and $\mu$ tends asymptotically to its critical value (but 
it still does not reach it until $T = 0$). This means that the 
$\tilde{S}[\bar{\varphi}]$ term in $\Gamma$ is always zero : there is no 
symmetry breaking on the Einstein universe and hence BEC does not occur.

\section{Finite Volume Spaces}
\label{sec:three}

In the previous section we examined numerically the Einstein static universe and 
showed that the critical temperature is zero, in contrast to calculations based 
on high temperature expansions (which implicitly assume that there is a non zero 
critical temperature!). We now wish to consider the generic case of a compact 
spatial manifold.

Usually in statistical mechanics, we are interested in the limit as both the 
volume and the number of particles tends to infinity. It is then meaningless to 
talk about $Q$, because if we imagine starting off at some low temperature with 
no anti-particles the charge will also become infinite. Thus we now restrict our 
attention to $\rho = Q/V$. For a general finite volume space, this will be given 
by (see appendix.\ref{appendix1} or \cite{HWPRD})
\begin{equation}
\rho = {2 \sinh (\beta \mu) \over V} \sum_{n=1}^{\infty} {n^2 e^{\beta \omega_n} 
\over \bigl( e^{\beta (\omega_n + \mu)} - 1 \bigr) \bigl( e^{\beta (\omega_n - 
\mu)} - 1 \bigr) }
\end{equation} 
where the $\omega_n$ satisfy $0 < \omega_1 < \omega_2 \leq \omega_3 \leq 
\ldots$~\  but are otherwise arbitrary.

The form of $\rho$ vs $\mu$ must be similar to fig.1: $\rho \to \infty$ whenever 
$\mu \to \omega_n$ and by similar arguments to those in section \ref{sec:one} 
$\mu$ lies in the region $| \mu | < \omega_1$ (the summand behaves like $\exp 
-\beta \omega_n$ as $\omega_n$ becomes large - which we expect to happen as $n 
\to \infty$ - therefore the sum is absolutely convergent). Thus we separate off 
the ground state contribution from $\rho$:
\begin{equation}
\rho = {2 \sinh (\beta \mu) e^{\beta \omega_1} \over \bigl( e^{\beta (\omega_1 + 
\mu)} - 1 \bigr) \bigl( e^{\beta (\omega_1 - \mu)} - 1 \bigr) } + \rho_e(\beta, 
\mu) .
\label{4.1}
\end{equation} and we consider the situation when $\mu$ is close to its critical
value. We will define $\mu = \omega_1 - \delta$, $\delta << 1$. $\rho_e(\beta, 
\mu)$ can be interpreted as the charge density in excited states.

Here (\ref{4.1}) is to be considered as an equation in the unknown 
$\mu$; $\rho$ is a measurable quantity which here we consider as fixed. 
Furthermore,
 all the equations which were used to derive (\ref{4.1}) are well 
defined at every point except the critical value (ie are well defined for all 
$\delta \not= 0$); thus no matter how close to the critical point we are, 
we must always be able to solve for $\mu$. But this is just the statement that 
the limit of the r.h.s of (\ref{4.1}) exists as $\delta \to 0$ and is equal 
to $\rho$. Now $\rho_e(\beta_c,\mu)$ tends to some limit, $\rho_{el}$ say, as 
$\delta \to 0$.(This must be true because the only divergent part is in the
ground state; the  sum in $\rho_e$ is absolutely convergent and therefore must
sum to some value at 
$\mu = \omega_1$). Let us call the ground state contribution in (\ref{4.1}) 
$\rho_g(\beta,\mu)$ and let us also define the constant $\rho_l = \rho - 
\rho_{el}$.

Assume that we have a critical point, ie a point where $\mu \to \omega_1$ as 
$\beta \to \beta_c$. Thus we must have that 
\begin{equation}
\rho_g(\beta, \omega_1 - \delta) \to \rho_l 
\quad
\hbox{as $\delta \to 0$.}
\end{equation} But by expanding $\rho_g$ in terms of $\delta$, it is easy to 
show
that
\begin{equation}
\rho_g \sim \frac{1}{\beta \delta V} + O\left(\frac{1}{V}\right)
\quad
\hbox{as $\delta \to 0$}
\end{equation} The only way that we can get a finite limit for $\rho$ as $\delta
\to 0$ (and 
$\beta \to \beta_c$) is if $\beta V = O(1/\delta)$ as $\delta \to 0$.
 
For a finite volume space we must have $\beta = O(1/\delta)$ as~$\delta \to 0$;
 thus $\beta \to 0$ as $\mu \to \omega_1$ 
 and the critical temperature is zero. Since for all finite $\beta$, 
$\mu \not= \omega_1$ this also means that there can be no symmetry breaking in 
such spaces. However, as we saw in the Einstein universe, if the volume is large 
we expect that $\mu$ can be quite close to its critical value for some finite 
temperature; thus there will be a large number of particles in the ground state 
and the physics will look similar to the infinite volume physics. In this 
situation the temperature at which we expect some degree of condensation is at  
a
critical temperature obtained by taking the infinite volume limit.  
Alternatively
one can try to define the onset of condensation by looking for the  maximum of
the specific heat since this is the property which is usually  searched for
experimentally.

\section{Conclusion}

We examined the Einstein static universe numerically, and showed that the 
critical temperature defined by $\mu \to \omega_1$ is zero in contrary to 
previous calculations. These gave the wrong result because they were based on 
high temperature expansions. We then generalized our result to arbitrary curved 
spaces of finite volume and showed that there is no symmetry breaking on these 
spaces at finite temperature and density.

\acknowledgments

JDS would like to acknowledge the financial support of the EPSRC.

\appendix
\section{Non interacting scalar field at finite temperature}
\label{appendix1}

In this appendix we wish to show in detail how the generalised zeta function 
techniques reproduce the earlier results of \cite{HWPRD}. Taking the 
field theory of section. \ref{sec:two} we define $\{\varphi_n\}$ s.t.
\begin{equation}
\bigl( - \nabla^2 + m^2 + U_1(x) \bigr)\varphi_n = \varepsilon^2_n \varphi_n.
\label{A1}
\end{equation} 
If we let $a = 2 \pi /\beta$ then
\begin{equation}
\zeta_{\cal G}(s) = \sum_n \sum_{j=-\infty}^{\infty} \left[ (aj + i\mu)^2 + 
\varepsilon^2 \right]^{-s}
\label{A2}
\end{equation} 
where the sum over $n$ in (\ref{A2}) is over all the eigenvalues
of (\ref{A1}).  Using (\ref{3.3}) and the Jacobi identity we obtain
\begin{equation}
\zeta_{\cal G}(s) = \frac{\pi^{1 \over 2}}{a\Gamma (s)} \int_0^{\infty} dt\ t^{s 
- {3 \over 2}} \left(\sum_n \exp (- \varepsilon^2_n t) \right) \left( 1 + 2 
\sum_{j=1}^{\infty} \exp \left(-{j^2 \pi^2 \over a^2 t}\right) \cosh \left({2\pi 
j\mu \over a} 
\right)\right)
\label{A3}
\end{equation}
 The integral in (\ref{A3}) naturally splits up into two parts. In
the second  part, the integral is easily done using the Bessel function integral
(by a  change of variable $t \to \varepsilon_n^2 t$):
\begin{eqnarray}
\frac{2\pi^{1 \over 2}}{a\Gamma (s)} \sum_n \sum_{j=1}^{\infty} \int_0^{\infty} 
\!&dt& t^{s - {3 \over 2}} \exp \left( -\varepsilon_n^2 t - {j^2 \pi^2 \over a^2 
 t}
\right) \cosh \left({2 \pi j\mu \over a}\right) \nonumber \\
& =& \frac{4\pi^{1
\over 2}}{a\Gamma (s)} \sum_n \sum_{j=1}^{\infty} \cosh 
\left( {2\pi j\mu \over a} \right) \left({\pi j \over a}\right)^{s - {1 \over 
2}} 
\varepsilon_n^{{1 \over 2} - s} K_{-s + {1 \over 2}}\left({2\pi j\varepsilon_n 
\over a}\right).
\end{eqnarray} 
Differentiating this, then evaluating at $s=0$ we obtain
(remembering  (\ref{3.8}) )
\begin{eqnarray}
\frac{4\pi^{1 \over 2}}{a} \sum_n \sum_{j=1}^{\infty} \cosh \left( {2\pi j\mu 
\over a} \right)\left({a \over \pi j}\right)^{1 \over 2}&&\!\!\!\!\!\! 
\varepsilon_n^{1 \over 2} K_{1 \over 2}\left({2\pi j\varepsilon_n 
\over a}\right) \nonumber \\ &=& 2 \sum_n
\sum_{j=1}^{\infty} \frac{1}{j} \cosh \left({2 \pi j\mu \over  a}\right) \exp
\left(- {2\pi j \varepsilon_n \over a}\right) \nonumber \\
&=& - \sum_n \left\{
\ln \bigl(1 - e^{-\beta (\varepsilon_n + \mu)}\bigr) + \ln 
\bigl(1 - e^{-\beta (\varepsilon_n + \mu)}\bigr) \right\}
\end{eqnarray}
where we have used the fact that $\beta = 2\pi /a$ and $\ln (1-x)
=  -\sum_{n=1}^{\infty} x^n$.

The first integral in (\ref{A3}) is
\begin{eqnarray}
\frac{\pi^{1 \over 2}}{a\Gamma (s)} \sum_n \int_0^{\infty}dt\ t^{s - {3 \over 
2}} \exp \bigl( - \varepsilon_n^2 t \bigr)  &=&\frac{\pi^{1 \over 2}}{a}
\frac{\Gamma (s - {1 \over 2})}{\Gamma (s)} \sum_n 
\frac{1}{\varepsilon_n^{2s - 1}} \nonumber \\
&=&\frac{\pi^{1 \over 2}}{a}
\frac{\Gamma (s - {1 \over 2})}{\Gamma (s)} 
\zeta_{\varepsilon_n^2}(s - 1/2)
\end{eqnarray} 

We expect $\zeta_{\varepsilon_n^2}(s - 1/2)$ to be an analytic function of $s$
except for a few simple poles. This implies that it has a well defined Laurent
expansion about $s=0$, which we will write as
\begin{equation}
\zeta_{\varepsilon_n^2} \sim \frac{A_{-1}}{2s} + A_0 + 2A_1s + O(s^2).
\end{equation}
Note that $A_{-1}$ is the residue of the generalised zeta function, using as the 
eigenvalues $\varepsilon_n^2$, at the point $s=1/2$.
This expansion, together with (\ref{3.8}), immediately gives
\begin{equation}
\left. \pi^{1 \over 2} \frac{\Gamma(s-{1 \over 2})}{\Gamma(s)} 
\zeta_{\varepsilon_n}(s - 1/2) \right|_{s=0} = \pi^{1 \over 2}\Gamma\left(-{1 
\over 
2}\right) A_{-1} = - 2\pi A_{-1}.
\end{equation} 
Also 
\begin{eqnarray}
\left[ \pi^{1 \over 2} \frac{\Gamma \left(s - {1 \over 2}\right)}{\Gamma (s)} 
\zeta_{\varepsilon_n}(s - 1/2) \right]' &\sim&\pi^{1 \over 2} \Gamma\left(s - {1 
\over
2}\right) \psi \left(s - {1 
\over 2}\right)A_{-1} \nonumber \\
& & +\ \pi^{1 \over 2} \Gamma (s - {1 \over2}) (A_0 + \gamma) + O(s)  
\quad
\hbox{as $s \to 0$,}
\end{eqnarray} 
where $\psi(z) = \Gamma'(z)/\Gamma(z)$.Making use of $\psi(-1/2) =
-\gamma + 2(1  - \ln 2)$ and then substituting in (\ref{2.6}) (remembering that
the 
$\tilde{S}[\bar{\varphi}]$ term never appears - see section \ref{sec:three}) 
gives
\begin{eqnarray}
\Gamma = \beta A_{-1} \ln l^2 &+& \beta \{ (2 - 2\ln 2 -\gamma) A_{-1} + A_0 + 
\gamma \} \nonumber \\
&+& \sum_n \ln \left[ \left( 1 - e^{-\beta (\varepsilon_n
+ \mu)} \right) \left(  1 - e^{- \beta(\varepsilon_n - \mu)} \right) \right].
\end{eqnarray} 
By comparison with the result of Haber and Weldon
\cite{HWPRD}, or from  examination of (\ref{A1}), we note that the
${\varepsilon}$ are the energy eigenvalues and identify
\begin{equation}
\sum_n \varepsilon_n = A_{-1} \ln l^2 + \{ (2 - 2\ln 2 -\gamma) A_{-1} + A_0 + 
\gamma \}
\end{equation} 
which is essentially the (regularized) Casimir energy.

Now , because $\varepsilon_n^2 = \sigma_n + m^2$,
\begin{eqnarray}
\zeta_{\varepsilon_n} &= &\frac{1}{\Gamma (s)} \int_0^{\infty} dt\ t^{s - 1} 
e^{- t (\sigma_n + m^2)} \nonumber \\
&=& \frac{1}{\Gamma (s)} \int_0^{\infty}
dt\ t^{s - 1} e^{- m^2 t} \Theta (t) .
\label{A4}
\end{eqnarray} 
It may be noted that $\Theta (t)$ is tr$\ \exp[ -t (-\nabla^2 +
U_1)]$ which has  a known asymptotic expansion as $t \to 0$ namely,
\begin{equation}
\Theta(t) \simeq (4 \pi t)^{- D/2} \sum_{k = 0, 1/2, 1, \ldots}^{\infty} t^k 
\theta_k .
\end{equation} The $\theta_k$ are the same coefficients that are used in
\cite{TomsPRD}, and  depend only on the geometry and the conditions on the 
fields
at the boundaries  of $\Sigma$. (Typically they involve products of the 
curvature
invariants and the  masses of the fields on $\Sigma$.) $D$ is the dimension of
$\Sigma$ and this is  the only place in which it enters explicitly. Use of this
expansion in  (\ref{A4}) together with expanding the $\exp (-m^2t)$ allows one 
to
find
\begin{equation} A_{-1} = \left\{ {4\pi^{- D/2} \sum_{l=0}^{D/2} (-1)^{D/2 - l}
\theta_{1/2 + l}\,  m^{D - 2l} \hbox{ for D even} \atop 4\pi^{- D/2} 
\sum_{l=0}^{(D
+ 1)/2} (-1)^{(D+1)/2 - l} \theta_l \, m^{D + 1 -2l} 
\hbox{ for D odd} } \right.
\end{equation} 
Thus the residue of $\zeta_{\varepsilon_n}(0)$ exists and
 can be  written down in general. As claimed earlier, one can see that
it depends only on  the geometry of the space. Unfortunately the authors are
aware of no similar  method for calculating $A_0$ in general, and one is reduced
to working out the  analytic continuation of $\zeta_{\varepsilon_n}$ case by 
case.
However, since the statistical mechanical contribution can be written down 
explicitly, one only needs to focus on the zero temperature zeta functions.

\section{Antipodal Identification}
\label{appendix2}

A simple modification of the Einstein static universe is to  identify antipodal 
points on $S^3$. Two cases are possible: periodic identification and 
antiperiodic identification of the fields. In the case of periodic 
identification, all the odd modes of $-\nabla^2$ must vanish. Thus the 
generalised zeta function in this case is
\begin{equation}
\zeta_{\lambda_{j,N}} = \sum_{j= -\infty}^{\infty} \sum_{N=0}^{\infty} {(2N + 
2)^2 \over \left[ \left({2\pi j \over \beta} + i\mu \right)^2 + {(2N + 1)(2N + 
3) \over a^2} + {6\xi \over a^2} + m^2 \right]^s} .
\label{B1}
\end{equation}

At first sight, it appears that we shall have to repeat the entire calculation 
for the zeta function above. However if one makes the change of variable, $n = N 
+ 1$, and notes that $(2N + 1)(2N + 3) = (2n -1)(2n + 1) = 4n^2 - 1$, we can 
rewrite (\ref{B1}) as
\begin{eqnarray}
\zeta_{\lambda_{j,n}} &=& \sum_{j=-\infty}^{\infty} \sum_{n=1}^{\infty} {4 n^2 
\over \left[ \left({2\pi j \over \beta} + i\mu \right)^2 + {4n^2 -1 \over a^2} + 
{6\xi \over a^2} + m^2 \right]^s} \nonumber \\
&=& 4 \times
\sum_{j=-\infty}^{\infty} \sum_{n=1}^{\infty} {n^2 \over \left[ (aj  + ib)^2 +
\alpha n^2 + c \right]^s} ,
\end{eqnarray} 
where $a = 2\pi/\beta$, $\alpha = 4/a^2$, $b= \mu$ and $c= (6\xi -
1)/a^2$. This  is the zeta function that we had before, and
\begin{eqnarray}
\Gamma_e =&&\frac{1}{32} \beta a^3 c^2 \ln (l^2c) - \frac{3}{64} \beta a^3 c^2 
 \nonumber \\
&&- \beta a^3 c^2 \sum_{n=1}^{\infty}
\frac{1}{\bigl(\pi nac^{1 \over  2}\bigr)^2} 
\left\{ K_2\bigl(\pi nac^{1 \over 2}\bigr) - \pi nac^{1 \over 2} K_3\bigl(\pi 
nac^{1 \over 2}\bigr) \right\} \nonumber \\
&&+ 4\sum_{n=1}^{\infty} n^2 \ln
\left[ \left(1 - e^{- \beta (\omega_n + \mu)} 
\right) \left(1 - e^{- \beta (\omega_n - \mu)} \right) \right]
\label{B2}
\end{eqnarray}
$\omega_n = \bigl((2n/a)^2 + c\bigr)$ and $c$ is as before. The factor of four 
in 
$\zeta$ appears simply because the space has half the volume that it had 
originally .As the volume becomes large we expect the log term in $\Gamma$ to 
also  be directly proportional to the volume, which explains the four here (as
the sum  tends to an integral, the measure will give a term proportional to
$a^3$. But in  (\ref{B2}) $a \to a/2$ compared to our original effective action,
leaving us  with  an overall factor of two). The spacetime gives us an effective
thermodynamic  mass, 
$c$, identical to the value on $S^3$, however the detailed dependence of the 
sums  at small volume is now radically different.

Because the expression (\ref{3.2}) is an absolutely convergent double sum for 
all points at which it is defined, we can write
\begin{equation}
\zeta_{odd} = \zeta_{S^3} - \zeta_{even}
\end{equation} which will be true even after analytic continuation.

Previously, it has been remarked that there would be many interesting properties 
of  this space associated with the possibility of have a $x$ dependent charge 
density.  However, this charge density appeared through the
$\tilde{S}[\bar{\varphi}]$  term  which we have seen will be absent.

\newpage

\end{document}